\documentclass[11pt]{article}
\usepackage{moriond,epsfig}

\def\Journal#1#2#3#4{{#1} {\bf #2}, #3 (#4)}

\def\PRL{\em Phys. Rev. Lett.}

\def\be{\begin{equation}}
\def\ee{\end{equation}}
\def\bea{\begin{eqnarray}}
\def\eea{\end{eqnarray}}

\renewcommand{\Im}{\mathop{\mathrm{Im}}}

\begin{document}
\title{COMPARISON BETWEEN ONE- AND TWO-CHANNEL KONDO EFFECTS IN
  MESOSCOPIC SYSTEMS}

\author{A. Zawadowski}

\address{Department of Theoretical Physics and Research Group of the
  Hungarian Academy of Sciences, Technical University of Budapest,
  H-1521 Budapest, Hungary}
\address{Solid State and Optical Research Institute of the Hungarian
  Academy of Sciences, H-1525 Budapest, Hungary}

\maketitle

\abstracts{There are strong recent indications that the Kondo
  phenomena play important role in the electronic transport and
  dephasing mechanism for some of the experimentally investigated
  metallic samples. The spin and the orbital Kondo problem with
  two-level systems provide very similar behavior above the Kondo
  temperature. Below the Kondo temperature the behavior is drastically
  different as in the latter case the entropy remains non zero. In the
  following several results derived earlier by different methods are
  simply reproduced by the time-ordered diagram technique in the
  leading logarithmic approximation valid above the Kondo temperature.}

In the recent years the evidences have been fast growing that in many
cases the low energy or low temperature electron transport in metals cannot be
explained in terms of electron-electron and electron-phonon
interactions. The anomalous behaviors occur e.g. in the electron
dephasing rate~\cite{Mohanty} and mesoscopic
transport~\cite{Pothier,Ujsaghy}. The leading candidates to resolve those
problems are the Kondo phenomena due to dynamical impurities. The
one-channel Kondo problem (1CK) is dealing with magnetic impurities,
the two-channel one (2CK) may describe the interaction between the
electrons and the fast relaxing two-level systems (TLS). In the 1CK
problem the characteristic Kondo temperatures $T_K$ cover a very broad
temperature range, while in the 2CK case the zero bias anomalies
indicate the typical range around $5$K~\cite{Ujsaghy}. In the second case the
theoretical description of the Kondo temperature is still in
debate~\cite{Aleiner,ZawaZarand} but the electron-hole symmetry
breaking in the electronic band structure seems to explain the
observed value. 

While the high temperature Kondo phenomena ($T>T_K$) are not very
different in the 1CK and 2CK cases, in the low temperature range their
behaviors are drastically different. Lowering the temperature
the impurity dynamics is largely enhanced around $T_K$. In the 1CK the
ground state is singlet, thus the dynamics is gradually frozen out and
Fermi liquid behavior is developing as the zero temperature is
approached. On the other hand the 2CK ground state without low energy
cutoff (e.g. level splitting) has a finite entropy, thus the enhanced
dynamics can exist even at very low temperature. In the 1CK case the
electron dephasing rate goes to zero as $T\to 0$, for the 2CK case a
new energy scale is suggested~\cite{ZDR99}, where the dephasing rate
is slowly varying even well below the Kondo temperature. While the
predictions are well confirmed for magnetic impurities, the role of
TLS is still subject of intensive discussions which is partially due
to the difficulties to identify the TLS-s in a real system. The non-universal
behavior of dephasing rate at low temperature supports that in some
cases the TLS may play a decisive role. 

The recent experiments carried out by the Saclay group~\cite{Pothier} 
in which the electronic transport in short wires is studied can be
also very different for different metals, but also are very sensitive
on impurities. The Ag and some of the Au samples show no anomalous
behavior, closely following the predictions based on electron-electron
interactions~\cite{Altshuler}. Some of the Au and the Cu samples show
a fast electronic energy relaxation by smearing the electron
distribution function around the Fermi energies corresponding to the
left and right contacts. Recently, it became known that some of the Au
samples contain fairly large amount of magnetic Fe
impurities~\cite{Pothier}. The
observed anomalies were explained in a phenomenological way by
assuming an anomalous electron-electron scattering rate which is
singular in the energy transfer E like $1/E^2$. 

Such a behavior first derived by J. Kroha~\cite{Kroha} in the
framework of the 2CK problem for $T<T_K$. Recently, it has been
pointed out by Kaminski and Glazman~\cite{Kaminski,Goppert} that a
similar behavior can be obtained for $T>T_K$ independently of the
number of the channels. The result for $T<T_K$ in the 1CK case must be
very different as in the framework of the Fermi liquid behavior no
singular behavior can be expected. 

The Kondo effect is due to the scattering of conduction electrons by a
localized object (magnetic or substitutional impurity or some defects)
with some internal degrees of freedom (e.g. spin, two close atomic
positions). The typical Hamiltonian of Kondo-like problems is
\begin{equation} 
H=\sum_{k\mu} \epsilon_k c^{\dagger}_{k\mu}c_{k\mu}  +
 \sum_{\alpha}\epsilon_\alpha b^{\dagger}_{\alpha}b_{\alpha} +
 \sum_{k,k'}\sum_{\mu\nu\alpha\beta} V_{\mu\nu}^{\alpha\beta}
 c^{\dagger}_{k\mu} c_{k'\nu} b^{\dagger}_{\alpha}b_{\beta}  
\label{eq:H}
\end{equation} 
where $\epsilon_k$ is the electron kinetic energy with momentum $k$,
$c^{\dagger}_{k\mu}$ creates an electron spherical wave with radial
momentum $k$ and internal quantum numbers $\mu$ and
$b^{\dagger}_{\alpha}$ creates a heavy object with quantum number
$\alpha$ ($\alpha$ being the spin, the position of the atom in the
TLS, or a crystal field
label of the impurity). Note that the internal indices $\mu$ and $\nu$
of the conduction electrons may also represent magnetic spin or
orbital indices or a combination of these as well.
$V_{\mu\nu}^{\alpha\beta}$ denotes the interaction coupling and a
band cutoff $D$ is applied.

That is obvious that a dynamical impurity with finite excitation
energy can mediate energy transfer between the electrons as a
localized phonon does. It was first pointed out in
Ref.s~\cite{Solyom} that magnetic impurities also mediate
energy transfer even in the absence of external magnetic field and
crystalline field splitting in third order of the exchange coupling,
while the second order contributes to the elastic
scattering~\cite{AGE}. 
\begin{figure}
  \begin{center}
    \epsfig{file=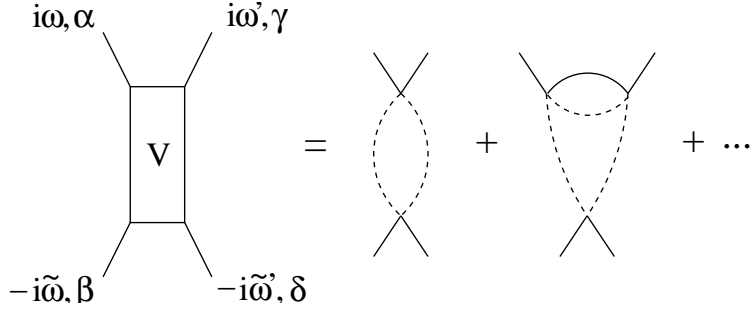,height=4cm}
    \caption{The effective magnetic impurity induced electron-electron
      interaction in the imaginary time technique. The diagrams
      contributing to lowest orders are also shown. The solid lines
      indicate the electrons, the dotted lines the Abrikosov's pseudofermions.
      \label{fig:1}}
  \end{center}
\end{figure}
The result in the lowest orders of the
perturbation for the electron-electron effective coupling for an
electron pair of zero total energy ($\omega=\tilde\omega$) shown in
Fig.~\ref{fig:1} is
\begin{equation}
  V_{\alpha\beta\gamma\delta}(\omega,\omega') =
  [V_{el}(\omega,\omega') +V_{inel}(\omega,\omega') ]
  \mbox{\boldmath{$\sigma$}}_{\alpha\gamma}
  \mbox{\boldmath{$\sigma$}}_{\beta\delta},
\end{equation}
where the elastic and the inelastic parts are respectively
\begin{equation}
  V_{el}(\omega,\omega') = \frac1T\frac13 S(S+1) J^2 \left[1+4J \rho_0
  \ln\frac D{|\omega|}\right] \delta_{\omega\omega'}
\end{equation}
and
\begin{equation}
  V_{inel}(\omega,\omega') = \frac13 S(S+1) J^2 4 (J\rho_0)
  \frac\pi{\omega-\omega'}
  (\mathop{\mbox{sgn}}\omega-\mathop{\mbox{sgn}}\omega'), 
\end{equation}
where $T$ is the temperature and $i(\omega-\omega')$ is the energy
transfer in which the inelastic part is singular. That vertex was used
to calculate the correction to the superconducting temperature in the
presence of magnetic impurities~\cite{Solyom}. In the analytic
continuation to real energy variables these contribute to the real
part of the $V_{inel}$ which is singular in the energy transfer $E$
($i(\omega-\omega')\to E$) as $1/E$.  

Kaminski and Glazman~\cite{Kaminski} were the first to point out that
the magnetic impurity induced electron-electron scattering rate is
also singular in the energy transfer. In the fourth order of the
perturbation theory that was proportional to $1/E^2$, as the energy
dependence assumed by the Saclay group~\cite{Pothier}.

In the following the derivation of that result with logarithmic
correction is presented in the non-vanishing leading logarithmic order.
The Abrikosov's pseudofermion technique~\cite{Abrikosov} is used,
where the $b$ operators in the Hamiltonian (\ref{eq:H}) create spin
states and $V_{\mu\nu}^{\alpha\beta}\,\Longrightarrow\,J
\mbox{\boldmath{$\sigma$}}_{\mu\nu}{\bf S}_{\alpha\beta}$ where $J$ is
the exchange coupling, $\mbox{\boldmath{$\sigma$}}$ is the Puli
operator and ${\bf S}$ is the spin operator describing the localized
spin. For simplicity the time ordered diagrams are used with real
time. That technique can be applied starting with arbitrary states
(e.g. non-equilibrium) and the transition amplitude between states
$|i\rangle$ and $\langle f|$ is given as
\begin{equation}
  \langle f|T|i\rangle_\omega = \sum_{n=1}^\infty \langle f|H_1
  \left(\frac 1 {\omega+i\delta -H_0} H_1 \right)^n |i\rangle,
\label{eq:fTi}
\end{equation}
where $\omega$ is the energy of the initial state. In the
Abrikosov's pseudofermion technique~\cite{Abrikosov} the decay rate is
calculated by taking the imaginary part of the forward scattering
amplitude (using the optical theorem) thus
\begin{equation}
  \frac1\tau \sim - \Im\{ \langle i|T|i\rangle_\omega\}.
\end{equation}
In that diagram technique the pseudofermion lines representing the spin
states (dotted line) are closed and there is only one line running
against the time insuring that at any moment only one spin state is present
for a given magnetic impurity. The electrons are represented by solid
lines.

\begin{figure}
\begin{center}
  \epsfig{file=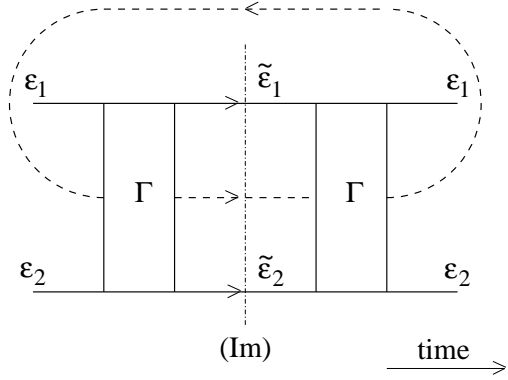,height=5cm}
  \caption{The two particle forward scattering amplitude $T(\omega)$
    giving the decay rate is shown in time ordered representation,
    where the intermediate state providing the imaginary part is also
    shown.
    \label{fig:2}}
\end{center}
\end{figure}
A typical forward scattering diagram which gives the result of Kaminski and
Glazman, e.g. in the electron-electron channel~\cite{Kaminski} in
the leading logarithmic order is shown in Fig.~\ref{fig:2}, where the
real part of vertex $\Gamma$ is taken and the cut (dashed dotted line)
represents the intermediate state where the imaginary part is
taken. 
\begin{figure}
  \begin{center}
    \epsfig{file=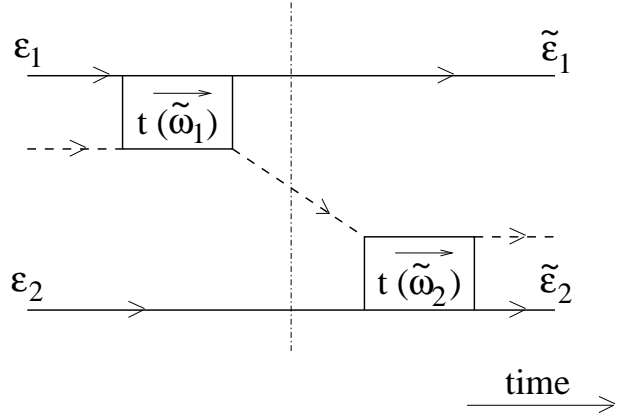,height=5.5cm}
    \caption{Typical time ordered diagram contributing to the vertex
      $\Gamma$ in the leading logarithmic approximation is shown. The
      square boxes represent the electron-pseudofermion vertexes
      $t(\tilde\omega)$ with total internal energy $\tilde\omega$.
      The dotted line represents the position of the energy denominator.
      \label{fig:3}}
  \end{center}
\end{figure}
In the logarithmic approximation the vertex $\Gamma$ is shown
in Fig.~\ref{fig:3}, where the square boxes are parquet vertex
correction $t(\tilde\omega)$ with energy $\tilde\omega$ running
through in the one particle channel and defined similarly to
Eq. (\ref{eq:fTi}). The dotted line inside the one particle scattering
amplitude runs in the direction of time.

Assuming that the total two-particle initial energy is $\omega$ and the
energy transfer is e.g. $E=\varepsilon_1-\tilde\varepsilon_1$, then the
intermediate energy denominator e.g. in the diagram in
Fig.~\ref{fig:3} is $\omega-\tilde\varepsilon_1-\varepsilon_2$, thus
with $\omega=\varepsilon_1+\varepsilon_2$ that is just the energy
transfer. In this way the two-particle interaction rate is $\Im T\sim
\frac1{E^2}$. 

The one-electron scattering blocks $t(\tilde\omega)$ are in
logarithmic approximation ($|\tilde\omega| + kT > T_K$)
\begin{equation}
  t(\tilde\omega) = \frac J{1-2J\rho_0 \ln \frac D{|\tilde\omega| +
  kT} }=\frac 1{2\rho_0 \ln \frac{|\tilde\omega| + kT}{T_K}}.
\end{equation}
where $T_K=D\exp\bigl \{-\frac 1{2 J\rho_0}\bigr \}$ is the Kondo
temperature in the leading logarithmic approximation~\cite{Gruner}.
In the diagrams in Fig.~\ref{fig:2} the one-particle vertex
corrections represented by the square boxes are separated in time, 
thus their internal
variables are $\tilde\omega_1=\omega-\varepsilon_1$ and
$\tilde\omega_2=\omega-\tilde\varepsilon_1$, thus with
$\omega=\varepsilon_1+\varepsilon_2$, $\tilde\omega_1=\varepsilon_1$
and $\tilde\omega_2 = \varepsilon_2+E$. In this way the scattering
rate is
\begin{equation}
  \frac1{\tau(\omega)} \biggr|_{\omega=\varepsilon_1+\varepsilon_2}
  \sim \frac1{E^2} \int (t(\varepsilon_1) t(\varepsilon_2+E))^2
  (1-n(\tilde \varepsilon_1))(1-n(\tilde \varepsilon_2))
  \delta(\omega- \tilde \varepsilon_1 -\tilde \varepsilon_2) d \tilde
  \varepsilon_1 d\tilde \varepsilon_2,
\label{eq8}
\end{equation}
where $n(\varepsilon)$ are the occupation numbers of the electronic
states in the actual non-equilibrium states. There are similar other
terms contributing to the right hand side of Eq. (\ref{eq8}), where
the vertex correction terms are of fourth order in the one-particle
vertices ($t(\varepsilon_1)$, $t(\varepsilon_2)$,
$t(\varepsilon_1+E)$, $t(\varepsilon_2-E)$) with different
combinations. That result is identical with Eq. (10) in Ref.~\cite{Kaminski}
only the arguments of the vertex corrections are somewhat different. 
It is important to point out, that the vertex correction term cannot
be factorized like $|\tau(\varepsilon_1)\tau(\varepsilon_2)|^2$ as it
was suggested in Ref.~\cite{Goppert}. See also Ref.~\cite{KrohaZawa}
where the slave boson technique is applied.

\begin{figure}
\begin{center}
\epsfig{figure=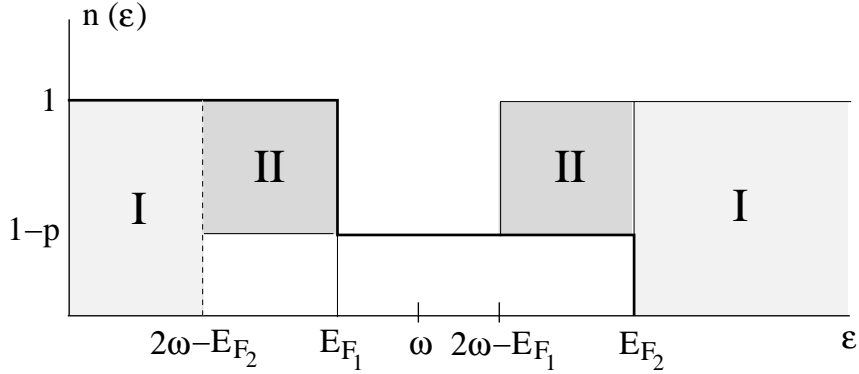,height=5cm}
\caption{The electron distribution function in a wire at large bias
  $e\,V$ is shown without the electron-electron interaction. The
  regions I and II are eliminated in the two steps, the heavy solid
  line separates the occupied and unoccupied regions.
\label{fig:4}}
\end{center}
\end{figure}

The vertex $t(\tilde\omega)$ can be easily calculated even for the
non-equilibrium case. Considering a short wire with bias $V$, the
Fermi energy for the contact 1(2) is denoted by $E_{F_1}(E_{F_1})$,
then $E_{F_2}-E_{F_1}= e\,V>0$. Assuming that the distance of the
impurity from contact 2 is $pL$ where $L$ is the length of the
sample and $0<p<1$. Then the non-interacting distribution of the
electrons at the impurity is represented by the heavy line in
Fig.~\ref{fig:4}. 
The scattering amplitude $t(\tilde\omega)$ can be
calculated by poor man's scaling by eliminating electron and hole states
above and below the energy $\tilde\omega$ in equal distances.

The elimination of states both occupied and unoccupied does not give
logarithmic contribution thus they do not contribute to the
scaling. The elimination is in two steps I and II. Take e.g. $\omega$
such that $E_{F_2}-\omega> \omega-E_{F_1}$, then in the first step I
the electron-hole states are eliminated for which $n=0$ or $n=1$. In
the step II only those electron states are eliminated which have
unoccupied counter hole states and $n\neq 0$, $n\neq 1$, thus only
portion $p$ of the
states. In the step II the initial coupling is the result obtained by
step I. Thus
\begin{equation}
  t^I(\tilde\omega) = \frac J{1-2J\rho_0 \ln \frac D{|E_{F_2}-
  \tilde\omega|} }
\end{equation}
and 
\begin{equation}
  t^{I+II}(\tilde\omega) = \frac {t^I(\tilde\omega)} {1-2\rho_0 p
  t^I(E_{F_2} -\tilde\omega) \ln \frac {|E_{F_2} -
  \tilde\omega|}{\tilde\omega -E_{F_1}}}.
\label{TIII}
\end{equation}
In Eq.~(\ref{TIII}) p occurs in the denominator. Here it is assumed
that $|E_{F_2}- \tilde\omega| > \tilde\omega -E_{F_1} > T_K^0$, where
$T_K^0$ is the bulk equilibrium Kondo temperature. In this case separate Kondo
resonances are formed~\cite{Kroha,KrohaZawa} at the two Fermi energies 
$E_{F_1}$ and $E_{F_2}$, and e.g. the width of the resonance described by the
separate Kondo temperature $T_K^1$ is defined by $\tilde\omega$ where
$t^{I+II}(\tilde\omega)\to\infty$. Thus
\begin{equation}
  T_K^1 = \frac{(T_K^0)^{\frac1p}} {(eV)^{\frac1p-1}}.
\end{equation}
That reproduces the result of Ref.~\cite{Coleman}. $T_K^2$ can be obtained by
electron-hole transformation and $p\to1-p$.

The inclusion of the Korringa type of spin life-time is beyond the
leading logarithmic approximation~\cite{ZawaPat}. In the case
of large bias $V$ the temperature is replaced by $eV$ and that can
play a dominating role which is not discussed here, but smears the $1/E^2$
singularity~\cite{Kaminski,Goppert,KrohaZawa} .

According to the experiments~\cite{Pothier} the scaling in the
distribution in the wire with a large bias holds only for the larger biases.
Above the Kondo temperature $n(E/eV)$ depends on only one
dimensionless variable as far as the Kondo corrections containing
$T_K^0$ (or in an other form $D$) are negligible. Recently J. Kroha
extended his slave boson calculation~\cite{Kroha} for $T>T_K$ and the
entire problem is treated in Ref.~\cite{KrohaZawa}.

\section*{Acknowledgments}

I express my gratitude to J. Kroha as the main part of the work has
been carried out in collaboration with him and I borrowed many of his
ideas. I am especially grateful among many others to H. Pothier,
N.O. Birge, M. Devoret, B. Altshuler, L.I. Glazman, J. von Delft,
O. \'Ujs\'aghy and G. Zar\'and for useful discussions.
This work was supported by Hungarian Grants OTKA T024005, T029813,
T034243 and by the A. v. Humboldt foundation.

\end{document}